\documentclass[
reprint,
 amsmath,amssymb,aps,prl,floatfix,
]{revtex4-1}

\usepackage{graphicx}% Include figure files
\usepackage{dcolumn}% Align table columns on decimal point
\usepackage{float}
\newcommand{\ie}{{\emph {i.e.}}}

\DeclareMathAlphabet{\bi}{OT1}{cmr}{bx}{it}
\newcommand{\br}{{\bi r}}
\newcommand{\Vdip}{{V_{\text{dip}}}}
\newcommand{\Vext}{{V_{\text{ext}}}}

\newcommand{\nmax}{n_{\rm max}}

\begin{document}

%\preprint{APS/123-QED}

\title{Candlestick Modes and Anisotropic Collapse of Dipolar Bose-Einstein Condensates}

\author{Jessica R. Taylor and Boaz Ilan}
\affiliation{School of Natural Sciences, University of California, Merced, 5200 North Lake Road, Merced, CA, 95343
}

%\pacs{L4-41: Nonlinear Dynamics}

\date{\today} 

\begin{abstract}
We use a gradient-decent method to compute 3D ground states of dipolar  Bose-Einstein condensates. We discover that in highly-prolate traps, whose long axis is parallel to the dipoles, can give rise to ``candlestick" ground states.
Direct numerical simulations of the dipolar Gross-Pitaevskii equation reveal that the nucleus of the candlestick mode undergoes collapse, while obtaining a highly flat pancake shape. The rate of this anisotropic collapse scales differently from what occurs in isotropic collapse. 
Stability analysis  reveals a surprising cusp point in the mass \emph{vs.} chemical potential curve, which may serve as a signature for this dynamics.

\end{abstract}


\maketitle

%\tableofcontents

Wave collapse is one of the hallmarks of nonlinear dispersive waves in a wide range of systems, including
plasma physics~\citep{robinson1997pa}, hydrodynamics~\citep{zeff2000singularity}, optics~\citep{moll2003self}, and Bose-Einstein condensates (BECs)~\citep{donley2001dynamics,lahaye2008d,wilson2009angular,eigen2016observation}.
Collapse occurs when an attracting (focusing) nonlinearity overcomes dispersion. This intrinsic feature of nonlinear waves has been investigated  for more than 50 years, leading to many theoretical advances in the framework of the nonlinear Schr\"odinger equation~\citep{sulem1999nonlinear,Fibich2015}. In general, it is found that collapse occurs with a self-similar profile~\citep{eggers2008role}. When the nonlinear interactions are local and isotropic, as is the case for the optical Kerr effect in bulk media and in some BECs, the wave collapses isotropically to a point. Radially-symmetric ring collapse in the isotropic NLS equation has also been studied and observed in optics~\citep{fibich2005new,vuong2006collapse,Grow:06,baruch2011singular}. 
BECs of $^{52}$Cr and other  atoms give rise to strong dipole-dipole interactions~\citep{Pethick2008}. Dipolar effects are also inherently anisotropic,  leading to anisotropic solitons~\citep{tikhonenkov2008anisotropic} and anisotropic superfluidity~\citep{ticknor2011anisotropic}.

Since BECs are prepared in energetic ground states, the characterization of these ground states is central to their dynamics. However, the theory of dipolar BEC ground states in three spatial dimensions (3D) is not well developed. Almost all the previous analytical and computational studies of dipolar BECs have relied on lower-dimensional approximations. In this Letter, we use an accurate numerical method to compute directly the 3D ground states of dipolar BECs trapped in anisotropic harmonic potentials. 
We find that in a highly-prolate trap,  the ground state has an elongated candlestick-like structure, with a larger nucleus flanked  by two smaller nodes.  Using direct simulations of the dipolar Gross-Pitaevskii equation, we show that the central node flattens and shrinks, eventually undergoing anisotropic collapse. We show that the collapsing nucleus deforms from an initially spherical shape to a pancake shape, which is retained during the collapse dynamics. The rate of this anisotropic collapse is found to scale  differently from the corresponding isotropic case. We also find  the mass \emph{vs.} chemical potential curve, which reveals a cusp point that has not been reported in other nonlinear systems. We raise the conjecture that such a cusp point is associated with anistropic collapse dynamics. 

We begin by recapitulating some of the key properties of dipolar BECs (see~\citep{Kevrekidis2007,Pethick2008,baranov2008theoretical,Lahaye2009}.).
The dynamics of a dipolar BEC can be described using  the dipolar Gross-Pitaevskii equation for the mean-field $\psi(\br,t)$,
\begin{align}
\label{eq:dGP}
i\hbar\psi_{t}(\br,t) &= \left[
-\frac{\hbar}{2m}\nabla^2 + \Vext(\br)
+ g_1\vert\psi(\br,t)\vert^2\right. \\[2mm]
& + \left. g_2\int_{\mathbb{R}^3} 
\Vdip(\br - \br')\vert\psi(\br',t)\vert^2 
d\br'\right] \psi(\br,t),
\nonumber
\end{align}
where  $\br = (x,y,z)$, 
$\nabla^2 = \partial^2_{xx} + \partial^2_{yy} + \partial^2_{zz}$ is the 3D Laplacian, 
$V_\text{ext}(\br)$ is an external potential, 
and the terms with $g_1$ and $g_2$ correspond to
the local (short-range) and dipolar (long-range) interactions, respectively.
The dipolar potential is 
\begin{equation}
\label{eq:Vdip}
\Vdip(\br) = \frac{1 - 3\cos^2\left(\theta\right)}{\vert\br\vert^3}
\end{equation}
where $\theta$ is the angle between $\br$ and the dipole axis, which we assume is aligned along the $z$-axis. We are interested a regime where the short-range interactions are repulsive, \ie, $g_1>0$, 
and the long-range interactions are such that $g_2<0$.
Due to the variation of the sign of~\eqref{eq:Vdip}, the dipolar interactions are give rise to both attractive and repulsive effects. Therefore, in the regime we consider, the dipolar interactions are the sole quasi-attractive mechanism.

Two important conserved quantities of Eq.~\eqref{eq:dGP} are the 
total number of atoms (analogous to the total power in optics),
\begin{align}
\label{eq:N}
N[\psi] = \int \vert\psi\vert^2d\br~,
\end{align}
and the total energy (Hamiltonian) of the condensate,
\begin{align}
\label{eq:E}
E[\psi] &= \frac{1}{2}\int_{\mathbb{R}^3} \left[\hbar^2\vert\nabla\psi\vert^2 + 
2V\vert\psi\vert^2 + g_1\vert\psi\vert^4\right. \\[2mm]
& + \left. g_2\int_{\mathbb{R}^3} \Vdip(\br - \br')\vert\psi(\br',t)\vert^2 
d\br'\vert\psi\vert^2\right]~.
\nonumber
\end{align}
BECs are formed in the energetic ground state, which can be characterized as follows. Assuming a time-harmonic solution, $\psi(\br,t) = u(\br)\exp(-i\mu t)$, where $\mu$ is called the chemical potential,  leads to the stationary dipolar GP equation for the mode structure function $u(\br)$,
\begin{align}
\label{eq:u}
\mu \hbar u(\br) &= \left[
-\frac{\hbar^2}{2m}\nabla^2 + \Vext(\br) + g_1\vert u(\br)\vert^{2} \right.  \\[2mm]
&+ \left. g_2\int_{\mathbb{R}^3} \Vdip(\br - \br')\vert u(\br')\vert^2 d\br'\right]u(\br). 
\end{align} 
Equation~\eqref{eq:u} admits infinitely many solutions. The ground state is the non-trivial 
minimizer of the total energy. 
It can be shown that the ground state is unique and non-negative~\citep{Lieb2001}. 
It is instructive and useful to reformulate Eq.~\eqref{eq:dGP} as (see~\citep{eberlein2005exact,Pethick2008})
\begin{subequations}
\label{eq:dGP-varphi}
 \begin{align}
i\hbar\psi_{t}(\br,t) &= \biggl[ 
-\frac{\hbar}{2m}\nabla^2 + \Vext(\br)  
+ (g_1-g_2)\vert\psi(\br,t)\vert^2  \\[2mm]
& + 3g_2  \varphi_{zz}(\br) \biggr] \psi~,
 \nonumber \\[2mm]
\nabla^2 \varphi(\br,t) &= |\psi(\br,t)|^2~,
\end{align}
\end{subequations}
where $\varphi(\br,t)$ is an auxiliary nonlinear potential that decays to zero as $|\br|\to\infty$. 
One advantage of this formulation is that it circumvents the highly-singular integral in~\eqref{eq:Vdip}.
 It is remarkable that system~\eqref{eq:dGP-varphi} has the same mathematical structure~\citep{footnote}
as the  Benney-Roskes  / Davey-Stewartson system, which describes  surface waves in a fluid of finite depth~\citep{benney1969wave,davey1974three} and intense light propagation coupled to an  electrostatic field~\citep{ablowitz1997multi,ablowitz2001nonlinear,crasovan2003arresting}. 
 This system has been studied extensively in (2+1)-dimensions. 
In particular,  in~\citep{ablowitz2005wave} it was shown that the collapsing solution becomes (mildly) anisotropic. However, almost all the previous studies have been in (2+1)-dimensions and not much is known about the solutions of this system in three spatial dimensions.

 Previous studies of dipolar BECs ground states
 have used variational methods to characterize the ground states, cf.~\citep{yi2001trapped,lushnikov2002collapse,ODell2004,eberlein2005exact,pedri2005two,bohn2009does,Lushnikov2010,abdullaev2013bright}. 
In the variational approach, one assumes a Gaussian or other profile for the shape of the mode and derives the relationships between its amplitude and width. However, as we show below, in certain parameter regimes
the structure of the ground state cannot be well-approximated with a Gaussian or any other simple analytic profile.
Here, we use a numerical approach to compute the ground states.

To fix some of the parameter regime,  we set  $\hbar=m=1$ 
and the  short and long-range interaction coefficients 
as $g_1=1$ and  $g_2=-1$. This is equivalent to normalizing the physical variables to the characteristic nonlinear 
length scale of the system, which is determined by the atomic scattering length.
We use an anisotropic harmonic external trapping potential, 
\begin{equation}
\label{eq:Vext}
\Vext(\br) = V_0 \left[ x^2 + y^2 + (\kappa z)^2\right]~,
\end{equation}
where $V_0$ is the potential depth and $\kappa$ is the anisotropy parameter, \ie, 
$\kappa < 1$ and $\kappa > 1$ correspond to 
a prolate (cigar-shaped) and oblate (pancake-shaped) potentials, respectively.
Since the trap is radially symmetric in the $x-y$ plane, the condensate inherits this symmetry.  
Thus, we denote the planar radial coordinate 
(not to the confused with $|\br| =  \sqrt{x^2+y^2+z^2}$) as
$$r_\bot=\sqrt{x^2+y^2}$$
and proceed to solve the stationary and time-dependent problems in the $(r_\bot,z)$ and 
$(r_\bot,z,t)$ coordinate systems, respectively.
To solve the stationary equation~\eqref{eq:u} 
we employ the Accelerated Imaginary-Time Evolution Method (A-ITEM), which is a gradient-decent method for computing ground states of nonlinear dispersive equations~\citep{Yang2008}. In particular, we iterate the A-ITEM scheme for Eq.~\eqref{eq:dGP-varphi} [recast in terms of $u(\br)$] while normalizing the peak density of the condensate, 
\begin{align}
\label{eq:amp}
\nmax = \max_{\br} |u(\br)|^2,
\end{align}
to a constant of choice. The chemical potential is recovered in each iteration from the relationship
\begin{align}
\label{eq:mu-E}
\mu = E[\psi] + \frac{1}{2} \left[\int_{\mathbb{R}^3} g_1\vert u\vert^4 \right.
+ \left.g_2\int_{\mathbb{R}^3} \Vdip(\br - \br')
\vert u(\br',t)\vert^2 d\br'\vert u\vert^2\right].
\end{align}
There is a one-to-one relation between  $\nmax$ and $\mu$. Therefore, the ground state can be defined by either of them. Convergence is obtained when the relative differences  between successive iterations of $\mu$ is smaller than  $10^{-5}$. In all cases, this requires at most a few thousand iterations.

\begin{figure}[ht!]
\centering
\includegraphics[width=\columnwidth]{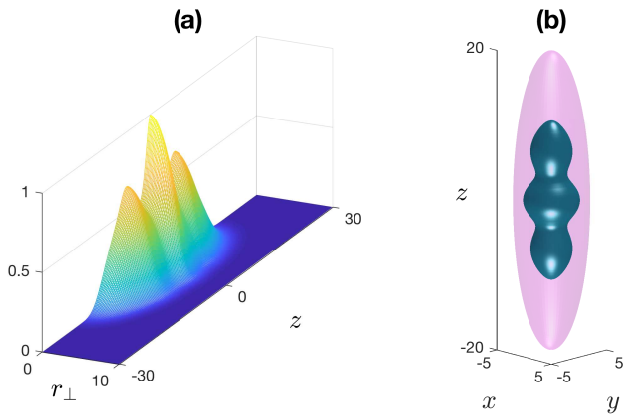}
\caption{\label{fig:iso_kappa=0_25} 
Candlestick condensate in a prolate trap with anisotropy parameter $\kappa = 4$ ,trap depth $V_0=0.1$, and condensate peak density $\nmax=1$.
(a) Density plot in the $r_\bot - z$ plane.
(b) Iso-surface plot of the condensate (internal, teal) surrounded by the trap (magenta). 
}
\end{figure}

We find that prolate traps can produce a ``candlestick"-shaped condensate, whose nucleus is flanked by two smaller nodes (see Fig. \ref{fig:iso_kappa=0_25}). In this figure we represent  the condensate in two different ways. First, as the density plot of $|\psi(\br)|^2 = u^2(\br)$ in the $r_\bot-z$ plane. Since the condensate is radially symmetric in the $x-y$ plane, the 3D ground state solution is computed by rotation of $u(r_\bot,z)$ around the $z$-axis. Fig. \ref{fig:iso_kappa=0_25}(b) presents an iso-surface plot of this 3D function, ``sliced" at the contour level of $0.4$, or $40\%$ of the peak density.
The candlestick modes arise when the trap is highly anisotropic and shallow. We emphasize that this  occurs in spite of the  trap having a single well, because of the anisotropic nature of the long-range interactions.

\begin{widetext}
\begin{figure}[ht!]
\centering
\includegraphics[width=2\columnwidth]{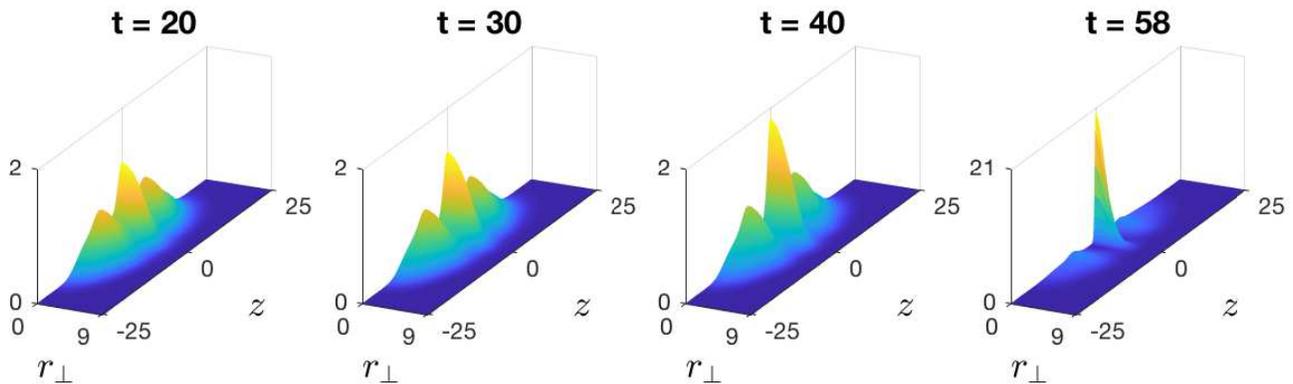}
\caption{\label{fig:candlestick_collapse_mesh2} 
Collapse of a candlestick condensate. Plots of $|\psi(r_\bot,z,t)|^2$ 
at different times. Note the greater range in the vertical axis in the last plot.
}
\end{figure}
 \end{widetext}

\begin{figure}[ht!]
\centering
\includegraphics[width=\columnwidth]{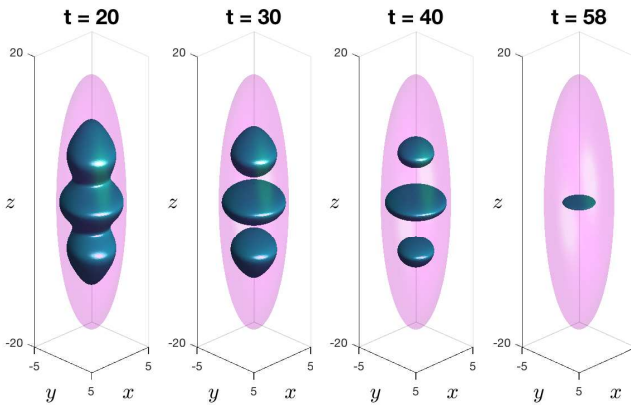}
\caption{\label{fig:candlestick_collapse_iso} 
Iso-surfaces corresponding to Fig.~.\ref{fig:candlestick_collapse_mesh2}.
}
\end{figure}

To investigate the evolution of the ground states, we solve Eq.~\eqref{eq:dGP-varphi}.
 Figures~\ref{fig:candlestick_collapse_mesh2} and~\ref{fig:candlestick_collapse_iso}
present four snapshots of the density and iso-surface dynamics of the solution. The initial excitation for this computation was the condensate in Fig.~\ref{fig:iso_kappa=0_25} with a $2\%$ amplitude perturbation, \ie,  $\psi(0,t) = 1.02\, u(\br)$.
We note that the collapse pattern in Fig.~\ref{fig:iso_kappa=0_25}(b) resembles qualitatively the collapse patterns observed experimentally by Metz \emph{et al.}~\citep{metz2009coherent} in $^{52}$Cr BECs. 
The nucleus might appear to break off from the two nodes. However, the nodes remain in tact and close to the nucleus. We point out that the two nodes do not appear in the last iso-surface plot, because their peak densities are much smaller than the peak density of the nucleus.
  
\begin{figure}[h!]
\centering
\includegraphics[width=\columnwidth]{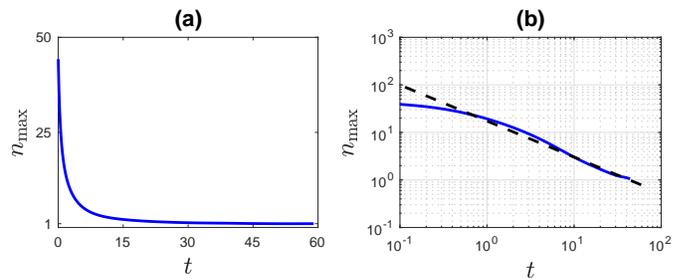}
\caption{\label{fig:blowup_and_loglog} 
The peak density of the collapsing nucleus using a  
normal plot (a) and a log-log plot (b) with a best power-law fit [Eq.~\eqref{eq:nmax-time}, dashed line]. 
}
\end{figure}

To monitor the collapse process of the condensate, we denote its peak density as
\begin{align}
\label{eq:nmax}
\nmax(t) = \max_{\br} |\psi(\br,t)|^2~.
\end{align} 
Figure~\ref{fig:blowup_and_loglog}(a) shows  $\nmax(t)$ for the collapse in Fig.~\ref{fig:candlestick_collapse_mesh2}, which blows-up at $T_c\approx 59\,$. Here $\nmax(0)=1$, so
$\nmax(t)$ is also the relative change of the peak density.
Figure~\ref{fig:blowup_and_loglog}(b) shows that, near the collapse time, the peak density scales as 
\begin{align}
\label{eq:nmax-time}
\nmax(t) \sim (T_c-t)^{-3/4}~.
\end{align}

\begin{figure}[h!]
\centering
\includegraphics[width=\columnwidth]{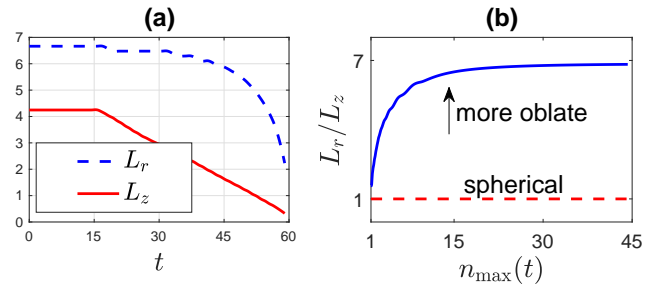}
\caption{\label{fig:aspect_ratio} 
The nucleus' (a) radial width (dashes) and thickness (solid) as functions of time. (b) The nucleus' aspect ratio \emph{vs.} its peak intensity during the collapse.
}
\end{figure}

 Figure~\ref{fig:aspect_ratio}(a) shows the radial width and thickness of the nucleus, which are recovered from the full-width at half-max of $|\psi(r_\bot,z,t)|^2$ along the radial and $z$-axes, respectively. After an initial stage $(t\le 15)$, the thickness $L_z(t)$ decreases linearly with $T_c-t$. Therefore, the peak density scales with the thickness  approximately as 
 \begin{align}
 \label{eq:nmax-thickness}
	\nmax(t) \sim [L_z(t)]^{-3/4}.
\end{align}
Figure~\ref{fig:aspect_ratio}(a)  also shows that the nucleus, which is initially almost spherical, flattens  more quickly than it shrinks in the radial directions, leading to a pancake-shaped collapse. 
This is further demonstrated in Fig.~\ref{fig:aspect_ratio}(b), which shows the aspect ratio of the nucleus plotted against its peak density during the collapse. The limiting aspect ratio near the collapse time is greater than 7, \ie, the pancake is quite flat, as also indicated by Figs.~\ref{fig:candlestick_collapse_mesh2} and~\ref{fig:candlestick_collapse_iso}.  In contrast, for  the analogous system in (2+1)-dimensions, the collapsing solution is mildly anisotropic~\citep{ablowitz2005wave}.

It is interesting to compare these results with the collapse in isotropic NLS equations. In NLS theory, the collapse (blowup) rate has has been studied extensively~\citep{sulem1999nonlinear,Fibich2015}.
Both the short-range and long-range nonlinearities in the dipolar GP equation~\eqref{eq:dGP} are cubic. For the 3D cubic NLS equation, it has long been conjectured and recently proven~\citep{merle2010stable} that, as the solution approaches the collapse time, its width decreases as $L\sim \sqrt{T_c-t}$, and its peak density increases as $\nmax \sim (T_c-t)^{-1}$. Therefore, $\nmax \sim 1/L^2$.
These exponents are quite different from~\eqref{eq:nmax-time} and~\eqref{eq:nmax-thickness}, further demonstrating the unique nature of anisotropic collapse.

\begin{figure}[ht!]
\centering
\includegraphics[width=\columnwidth]{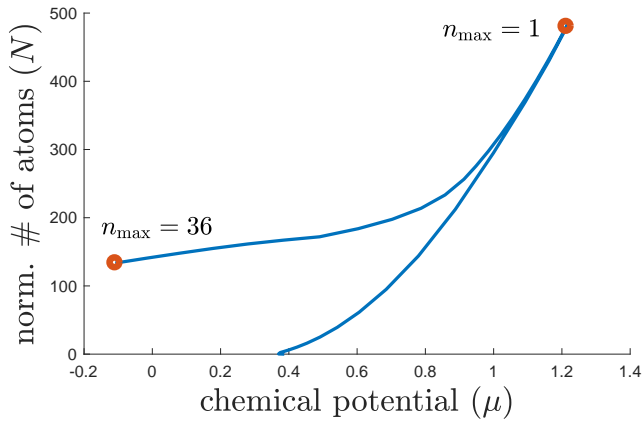}
\caption{\label{fig:plot_slope_candlestick} Normalized number of atoms of candlestick condensates as the chemical potential is varied.
}
\end{figure}

We also study the linear stability of these ground states. Figure~\ref{fig:plot_slope_candlestick} shows the (normalized) total number of atoms~\eqref{eq:N} in the ground state as functions of the chemical potential. The potential depth is $V_0=0.1$ and anisotropy parameter is $\kappa=0.25\, $. This graph is obtained by computing the ground states with peak densities in the range   $0.1 \le \nmax  \le 36$ and using Eq.~\eqref{eq:mu-E} to map $\nmax$ to $\mu$.
In NLS theory, it is known that the ground state is  amplitude-stable when  $dN / d\mu >0$~, where $N(\mu) = N[\psi(\cdot;\mu)]$ is the number of atoms~\eqref{eq:N} or total mass~\citep{vakhitov1973stationary,sivan2008qualitative,ilan2011quantitative}.
In this sense, Fig.~\ref{fig:plot_slope_candlestick}  indicates that the ground states are linearly stable, 
despite their nonlinear instability. However, what is truly surprising about Fig.~\ref{fig:plot_slope_candlestick} is the cusp point, which occurs at $\nmax \approx 1$. To our knowledge, this is the first system in which a cusp point in $N(\mu)$ has been reported.  Moreover, we note that, as $\nmax$ increases, \ie, as one traces the ground states along the $N(\mu)$ curve, the nucleus becomes much larger compared with the side nodes. This mimics what happens during the collapse dynamics. Based on this observations, we conjecture that the cusp point in the $N(\mu)$ curve could be a signature of the anisotropic collapse of these ground states.

In conclusion, our computational results show that  long-range dipolar interactions in a BEC can give rise to highly anisotropic condensate ground states that undergo anisotropic collapse, which occurs at a different rate from the isotropic case. 
This demonstrates that anisotropic long-range interactions can lead to localized nonlinear waves with unique collapse patterns and kinetic properties, which could open up new venues  for exploring many-body collective excitations.

\begin{acknowledgements}
The authors gratefully acknowledge computing time on the Multi-Environment Computer for Exploration and Discovery (MERCED) cluster at UC Merced, which was funded by National Science Foundation Grant No. ACI-1429783.
\end{acknowledgements}

\newpage

% \bibliography{GSmanCit}

\begin{thebibliography}{42}
\bibitem [{foo()}]{footnote}%
  \BibitemOpen
  \href@noop {} {}\bibinfo {note} {The Benney-Roskes / Davey-Stewartson system
  can be obtained from Eq.~\eqref{eq:dGP-varphi} by assuming that $\psi(\br,t)$
  does not depend on $x$, $\Vext=0$, and making the transformation $z\to x,
  y\to y$ and $\phi = \varphi_x\, $.}\BibitemShut {Stop}%
\end{thebibliography}

%

\end{document}